\documentclass[conference]{IEEEtran}
\IEEEoverridecommandlockouts
\usepackage{cite}
\usepackage{amsmath,amssymb,amsfonts}
\usepackage{algorithmic}
\usepackage{graphicx}
\usepackage{textcomp}
\usepackage{xcolor}
\DeclareMathOperator*{\argmin}{argmin}
\def\BibTeX{{\rm B\kern-.05em{\sc i\kern-.025em b}\kern-.08em
    T\kern-.1667em\lower.7ex\hbox{E}\kern-.125emX}}
\begin{document}

\title{Survey Paper on Control Barrier Functions}

\author{\IEEEauthorblockN{Promit Panja}
\IEEEauthorblockA{\textit{Dept. of Electrical and Computer Engineering} \\
\textit{Virginia Tech}\\
Blacksburg, United States \\
ppanja@vt.edu}
}

\maketitle

\begin{abstract}
Control Barrier Functions (CBFs) have emerged as a powerful paradigm in control theory, providing a principled approach to enforcing safety-critical constraints in dynamic systems. This survey paper comprehensively explores the foundational principles of CBFs, delves into the complexities of High Order Control Barrier Functions (HOCBFs), and extends the discussion to the adaptive realm with adaptive Control Barrier Functions (aCBFs). Through a systematic examination of theoretical underpinnings, practical applications, and the evolving landscape of research, this survey highlights the versatility of CBFs in addressing safety and stability challenges.
\end{abstract}


\section{Introduction}
\textit{Cyber-Physical Systems} couple control, computation, and physical dynamics into one integrated system, and when safety is a major design consideration they are considered as \textit{safety-critical} \cite{ames2016control, ames2019control}. Before we talk about ensuring safety we first need to answer the question of what is \textit{safety}? The authors in \cite{ames2019control} provide a neat answer to this question, safety requires that “bad” things do not happen while liveness requires that “good” things eventually happen, e.g., asymptotic stability can be seen as an example of a liveness property in the sense that an asymptotically stable equilibrium point is eventually reached. Conversely, invariance can be seen as an example of a safety property in the sense that any trajectory starting inside an invariant set will never reach the complement of the set, describing the locus where bad things happen. The requirement for safety is amplified by the increasing complexity in modern CPS. As these systems become integral to critical infrastructures and domains such as automotive, aerospace, industrial automation, and robotics the potential consequences of safety lapses escalate. Consequently, a comprehensive understanding of safety properties, encompassing not only classical notions like stability but also embracing advanced concepts in control theory, becomes paramount.

The concept of safety in dynamical systems is nothing new. In the past multiple research studies have been conducted in the field of \textit{safe-autonomy} but formally necessary and sufficient conditions for safety in terms of set invariance were provided by Nagumo in the 1940's \cite{nagumo1942lage}.

Given a dynamical system,
\begin{equation}
    \dot{x} = f(x)
\end{equation}

where $x \in \mathbb{R}^{n}$, assuming that a safe set $\mathcal{C}$ is a superlevel set of a smooth function $h : \mathbb{R}^{n} \to \mathbb{R}$, i.e., $\mathcal{C} = \{x \in \mathbb{R}^{n} : h(x) \geq 0\}$, and that $\frac{\partial h}{\partial x}(x) \neq 0$ for all $x$ such that $h(x) = 0$, then Nagumo's theorem gives sufficient and necessary conditions for set invariance for the derivative of $h$ in the boundary of $\mathcal{C}$ \cite{ames2019control,xiao2023safe}:

\begin{equation}
    \mathcal{C}\ \text{is invariant} \iff \dot{h}(x) \geq 0 \ \forall x \in \partial\mathcal{C}
\end{equation}

Similarly, \textit{barrier certificates} were introduced as an important tool for formally verifying the safety of nonlinear and hybrid systems, the term "barrier" was chosen from optimization where barrier functions are added to cost functions to avoid undesirable regions. Barrier certificates seem to discover Nagumo's conditions independently \cite{ames2016control,ames2019control,xiao2023safe}. In this case one considers an unsafe set $\mathcal{C}_{u}$ and a set of initial conditions $\mathcal{C}_{0}$ together with a function $B : \mathbb{R}^{n} \to \mathbb{R}$ where $B(x) \leq 0$ for all $x \in \mathcal{C}_{u}$ and $B(x) > 0$ for all $x \in \mathcal{C}_{u}$. Then $B$ is a barrier certificate if:

\begin{equation}
    \dot{B}(x)\leq 0 \implies \mathcal{C}\ \text{is invariant}
\end{equation}

Safety can be motivated by analyzing the stability of systems. Let us assume we need to \textit{asymptotically} stabilize a nonlinear control system at $x^{*} = 0$. In a nonlinear context, this can be achieved by a feedback control law that drives a positive definite function, $V : D \subset \mathbb{R}^{n} \to \mathbb{R}_{\geq0}$, to zero. That is if $\exists u = k(x) \text{s.t.} \dot{V}(x, k(x)) \leq -\gamma(V(x))$, where $\dot{V}(x, k(x)) = L_{f}V(x)+ L_{g}V(x)k(x)$ then the system is stabilized to $V(x^{*}) = 0$. Here $\gamma$ is a class $\mathcal{K}$ function, defined on the real number line, maps from $\mathbb{R}_{\geq0} \to \mathbb{R}_{\geq0}$, and is strictly monotonic. $V$ is a \textit{Control Lyapunov Funciton (CLF)} if it is positive definite and it satisfies \cite{ames2019control,xiao2023safe}:

\begin{equation}
    sup_{u \in U}[L_{f}V(x) + L_{g}V(x)u] \leq - \gamma (V(x))
\end{equation}

$\gamma$ is a class $\mathcal{K}$ funciton, such that we can consider a set of all stabilizing controllers for every point $x \in \mathcal{D}$:

\begin{equation}
    K_{clf} := \{u \in U : L_{f}V(x) + L_{g}V(x)u \leq - \gamma (V(x))\}
\end{equation}

This survey aims to provide a comprehensive overview of the current state-of-the-art Control Barrier Functions, shedding light on their theoretical foundations, practical applications, and the evolving landscape of research in this domain. In this paper we are going to cover mainly three types--- regular Control Barrier Function (CBF), High-Order Control Barrier Function (HOCBF), and Adaptive Control Barrier Function (aCBF).

The remainder of this paper is organized as follows. Section II covers the math preliminary and theory of the previously mentioned three types of CBFs. In Section III we look at some of the applications of CBFs and study their implementations. We close our discussion with our conclusion in Section IV and discuss some of the future work in Section V.

\section{Theory}

In this section, we first cover the required mathematical concepts like safe set and set invariance, then formally introduce \textit{Control Barrier Functions (CBFs)}. We then move forward to study two variations of CBFs, \textit{Higher-Order Control Barrier Functions (HOCBFs)} and \textit{Adaptive Control Barrier Functions (aCBFs)}. 

\subsection{Control Barrier Functions (CBFs)}

Suppose a nonlinear affine control system is of the form:
\begin{equation}
    \dot{x} = f(x) + g(x)u
\end{equation}

Where $x \in \mathbb{R}^{n}$ is the system state, and $u \in \mathbb{U} \subseteq \mathbb{R}^{m} $ is the control input. The functions $f: \mathbb{R}^{n} \to \mathbb{R}^{n}$ and $g: \mathbb{R}^{n} \to \mathbb{R}^{n \times m}$ are locally Lipschitz. Then there exists a unique solution $x(t)$ to equation (6) for an initial system state $x(t_{0}) = x_{0}$ and a maximum time interval $I(x_{0}) = [0, t_{max})$.

Assume a safe set $\mathcal{C}$ which contains all the states where the system is considered safe which is defined as:
\begin{align}
    \begin{split}
        \mathcal{C} ={}& \{x \in \mathbb{R}^{n}: h(x) \geq 0 \}
    \end{split} \\
    \begin{split}
        \partial\mathcal{C} ={}& \{x \in \mathbb{R}^{n}: h(x) = 0 \}
    \end{split} \\
    \begin{split}
        Int(\mathcal{C}) ={}& \{x \in \mathbb{R}^{n}: h(x) > 0 \}
    \end{split}
\end{align}

Where $\mathcal{C}$ is a superlevel set of the continuously differentiable function $h: \mathbb{R}^{n} \to \mathbb{R}$, $\mathcal{C}$ is non-empty and has no isolated points and $\frac{\partial h}{\partial x} \neq 0$. Then the set $\mathcal{C}$ is said to be forward invariant if for every initial state $x_{0} \in \mathcal{C}$, $x(t) \in \mathcal{C}$ for all $t \in I(x_{0})$ \cite{ames2019control}.

If the safe set $\mathcal{C}$ is proved to be forward invariant then the solution to (6) would always remain inside of this safe set $\mathcal{C}$. This means the system would always be in a safe state when starting from a given initial state inside the safe region. Therefore the system (2) is considered \textit{safe} with respect to the set $\mathcal{C}$ if the set $\mathcal{C}$ is forward invariant.

From our previous discussion of Control Lyapunov Functions, we would like to generalize the concept of safety for a system. If there exists a CLF $V$ such that $V(x) = 0 \implies x \in \mathcal{C}$ and $V$ has a superlevel set $\Omega_{c} = \{x \in D : v(x) \leq c\} \subset \mathcal{C}$, then the controller in (5) would render $\Omega_{c}$ invariant and set $\mathcal{C}$ safe. But it is overly restrictive which means it would render every sublevel set as invariant. We wish to enforce set invariance without requiring a positive function, for $h$ to be a barrier function it should render only $\mathcal{C}$ as invariant and not its sublevel sets \cite{ames2019control,ames2016control,xiao2023safe}.

Although barrier functions are a great tool to verify set invariance, they cannot be directly used to design controllers that ensure invariance and safety \cite{ames2016control}. Due to this reason, we formulate control barrier functions, similar to how Lyapunov functions are extended to control Lyapunov functions that ensure asymptotic stability.

{\bf Definition 1:} For a set $\mathcal{C}$ defined as (7)-(9) for a continuously differentiable function $h: \mathbb{R}^{n} \to \mathbb{R}$, the function $h$ is called control barrier function on set $\mathcal{D}$ with $\mathcal{C} \subseteq \mathcal{D} \subseteq \mathbb{R}^{n}$, if there exists a class $\mathcal{K}$ function $\alpha$ such that,
\begin{equation}
    sup_{u \in U}[\underbrace{L_{f}h(x) + L_{g}h(x)u}_\textrm{$\dot{h}(x, u)$}] \geq - \alpha (h(x)), \forall x \in \mathcal{D}
\end{equation}

Where, $\dot{h}(x, u) =\nabla h(x) \cdot \dot{x}$. $L_{f}h(x) = \nabla h(x) \cdot f(x)$ and $L_{g}h(x) = \nabla h(x) \cdot g(x)u$ are Lie derivative terms. $\alpha$ is known as a class $\mathcal{K}$ function if $\alpha : [0, \infty) \to [0, \infty)$ with $\alpha (0) = 0$ and must be continuously increasing.

Now that we have established that CBFs provide necessary and sufficient conditions for safety, we can now utilize this notion to build controllers. We want to build a controller or a \textit{safety filter} that minimally modifies a sample desired input $u_{des}$ that enforces the system to stay inside the safety region. This can be achieved by utilizing optimization-based controllers.
\begin{align}
    u^{*} = \argmin_{u \in \mathbb{U} \subseteq \mathbb{R}^{m}} \parallel u - u_{des}\parallel^{2} \\
    \textrm{s.t.} \quad L_{f}h(x) + L_{g}h(x)u \geq - \alpha h(x)
\end{align}

The Quadratic Program (QP) defined above finds the minimum perturbation in $u$. For a single inequality constraint without an input constraint, the QP has a closed-form solution as per the KKT conditions. This is called the Control Barrier Function based Quadratic Program (CBF-QP), by solving the above CBF-QP we can obtain an optimal value of $u^{*}$.

One thing that needs to be kept in mind is regarding the feasibility of the CBF-QP. CBF-based optimization problems might not have a feasible solution. This definitely would be true if the original problem to begin with is not feasible. Feasibility depends on the possible conflict between CBF constraints, original system constraints, and constraints in the control input. It also depends on the choice of the class $\mathcal{K}$ function that the control designer might select. Another reason might be due to discretization QP-based solutions. A wrong choice of $\Delta t$ might cause problems related to \textit{inter-sampling effect}, it needs to be made sure that CBF constraint is satisfied in every time interval or else the forward invariant property may be violated \cite{xiao2023safe}.

In the next section, we will discuss what happens when the control input does not show up explicitly in the first time-derivative of $h(x)$. This requires extending CBFs to higher orders and formulating \textit{High-Order CBF (HOCBF)}.

\subsection{High-Order Control Barrier Functions (HOCBFs)}

We can see that in order for a controller to exist that ensures the safety condition, the $L_{g}h(x) \neq 0$. This suggests that the control input $u$ must be present in the CBF constraint which is the first time-derivative of $h(x)$. One might quickly encounter problems in systems where the control input is not explicitly present. This limits the application of CBFs to only the systems with relative degree of one.

Many studies have been conducted to extend CBFs to high relative degrees. \cite{hsu2015control} introduced a backstepping approach for high relative degree constraints, and was shown to work for relative degrees of 2. \cite{wu2015safety} showcased a position-based method for relative degrees of 2. In \cite{nguyen2016exponential} a more general form was introduced that works for arbitrarily high relative degree constraints, by employing input-output linearization and pole placement controller with negative poles that stabilizes the CBF to zero. A simpler form of HOCBF was proposed in \cite{xiao2019control}.

As shown in \cite{ames2019control}, assume $h(x)$ has arbitrary high relative degree $r > 1$, we can write the $r^{th}$ time derivative as,

\begin{equation}
    h^{(r)}(x, u) = L^{r}_{f}h(x) + L_{g}L^{r-1}_{f}h(x)u
\end{equation}

where, $L_{g}L^{r-1}_{f}h(x) \neq 0$ and $L_{g}L_{f}h(x) = L_{g}L^{2}_{f}h(x) = \dotsm = L_{g}L^{r-2}_{f}h(x) = 0, \forall x \in \mathcal{D}$. 

Next, we can write,

\begin{equation}
    \eta_{b}(x) := \begin{bmatrix} h(x) \\ \dot{h}(x) \\ \ddot{h}(x) \\ \vdots \\ h^{(r-1)}(x) \end{bmatrix} = 
    \begin{bmatrix} h(x) \\ L_{f}h(x) \\ L^{2}_{f}h(x) \\ \vdots \\ L^{r-1}_{f}h(x) \end{bmatrix}
    \end{equation} 

Now from \cite{xiao2023safe}, for a $r^{th}$ order differentiable function as shown above $h(x) : \mathbb{R}^{n} \to \mathbb{R}$, we recursively define a sequence of functions $\psi_{i} : \mathbb{R}^{n} \to \mathbb{R}, i \in \{1, \dots, r\}$ in the form,

\begin{equation}
    \psi_{i}(x) = \dot{\psi}_{i-1}(x)+ \alpha_{i}(\psi_{i-1}(x)), i \in \{1, \dots, r\}
\end{equation}

where, $\alpha_{i}, i \in \{1, \dots, r\}$ is a set of class $\mathcal{K}$ functions. We then define a sequence of sets $\mathcal{C}_{i}, i \in \{1, \dots, r\}$ in the form,

\begin{equation}
    \mathcal{C}_{i} = \{x \in \mathbb{R}^{n} : \psi_{i-1}(x) \geq 0\}, i \in \{1, \dots, r\}
\end{equation}

The set $\mathcal{C}_{1} \cap, \dots, \cap \mathcal{C}_{r}$ is \textit{forward invariant} if $h(x)$ is a HOCBF.

{\bf Definition 2:} For $\mathcal{C}_{i}$ defined as (16) and $\psi_{i}(x)$ defined as (15), where $i \in \{1, \dots, r\}$ for a continuously differentiable function $h: \mathbb{R}^{n} \to \mathbb{R}$, the function $h$ is called high order control barrier function (HOCBF) of relative degree $r$, if there exists differentiable class $\mathcal{K}$ functions $\alpha_{i}, i \in \{1, \dots, r\}$ such that,
\begin{equation}
    sup_{u \in U}[L^{r}_{f}h(x) + L_{g}L^{r-1}_{f}h(x)u + O(h(x)) + \alpha_{r}(\psi_{r-1}(x))] \geq 0
\end{equation}

\begin{equation}
    \text{where,} \ O(h(x)) = \sum^{r-1}_{i=1} L^{i}_{f}(\alpha_{m-1} \circ \psi_{m-i-1})(x)
\end{equation}

for all $x \in \mathcal{C}_{1} \cap, \dots, \cap \mathcal{C}_{r}$. $L_{f} \text{and} L_{g}$ denote the partial Lie derivatives w.r.t $x \ \text{along} \ f \ \text{and} \ g$ \cite{xiao2023safe}.

Now, to formulate an HOCBF-based optimization problem that guarantees safety for systems with arbitrary relative degree, we replace the CBF constraint in (12) with the HOCBF constraint (17) \cite{xiao2023safe,ames2019control,xiao2021adaptive,liu2023auxiliary,unknown}.

\subsection{Adaptive Control Barrier Functions (aCBFs)}

Until now we have discussed CBFs and and their extension HOCBFs. We delved into the mathematical frameworks which underpin these methods. We saw that optimizing quadratic costs while stabilizing affine control systems to desired (sets of) states subject to state and control constraints can be reduced to a sequence of quadratic programs (QPs) by using control barrier functions (CBFs) and control Lyapunov functions (CLFs) \cite{xiao2021adaptive}.

While traditional CBFs and HOCBFs have exhibited remarkable efficacy in handling constrained systems with known dynamics, in the real-world, uncertainties and varying operating conditions can cause the CBF framework to break. These uncertainties can arise from factors such as parameter variations, external disturbances, or the inherent unpredictability of certain dynamic processes. As such, there is a growing need to equip control systems with the ability to adapt to these uncertainties in real-time, ensuring a robust and reliable performance across diverse scenarios.

One of the goals of adaptive control is for parameter variations. Parameters may vary due to changes in the operating environment, nonlinear actuators, and disturbances acting on the system \cite{aastrom2013adaptive}. Adaptive control Lyapunov functions (aCLFs) were introduced close to 20 years ago, and provide a solid foundation for Lypunov theory based methodology to design controllers that stabilize systems with parameter uncertainty. In this section, we will be using this concept in the context of safety to define a framework which adaptively achieves safety by keeping the system’s state within a safe set even in the presence of parametric model uncertainty \cite{taylor2020adaptive}.

As we transition into the realm of Adaptive Control Barrier Functions, it becomes apparent that the dynamism inherent in many real-world systems necessitates a more flexible and responsive approach. This section explores the integration of adaptability into the CBF framework, aptly named \textit{Adaptive Control Barrier Functions (aCBFs)}, as a solution to the challenges posed by dynamic and evolving environments.

We start by modifying the system shown in (6) to a system with uncertainties in dynamics,

\begin{equation}
    \mathbf{\dot{x}} = \mathbf{f}(\mathbf{x}) + \mathbf{F}(\mathbf{x})\mathbf{\theta^{*}} + \mathbf{g}(\mathbf{x})\mathbf{u},
\end{equation}

where, $\mathbf{\theta^{*}} \in \Theta \subset \mathbb{R}^{p}$ is a vector of unknown parameters. $\mathbf{F} : \mathcal{X} \to \mathbb{R}^{p \times q}$ is assumed to be smooth on $\mathcal{X}$ with $\mathbf{F}(\mathbf{0}) = \mathbf{0}$. Now we consider controllers that update an estimate of the unknown parameters of the form,
\begin{align}
    \begin{split}
            \mathbf{u} = \mathbf{k}(\mathbf{x}, \mathbf{\Hat{\theta}})
    \end{split} \\
    \begin{split}
            \mathbf{\dot{\Hat{\theta}}} = \mathbf{\Gamma} \mathbf{\tau}(\mathbf{x}, \mathbf{\Hat{\theta}})
    \end{split},
\end{align}

where, $\mathbf{\Hat{\theta}} \in \Theta$ represents the estimate of the parameters $\mathbf{\theta^{*}}$, maintained by the controller, $\mathbf{k} : \mathcal{X} \times \Theta \to \mathcal{U}$ is the control law, $\mathbf{\Gamma} \in \mathbb{R}^{n \times n}$ is the matrix adaptive gain, $\mathbf{\tau} : \mathcal{X} \times \Theta \to \mathbb{R}^{p}$ is the adaptation law with the assumptions, $\mathbf{k}$ is locally Lipschitz, $\mathbf{\tau}$ is locally Lipschitz, and $\mathbf{\Gamma}$ is symmetric and positive definite \cite{taylor2020adaptive,aastrom2013adaptive,xiao2021adaptive}.

After introducing the above parameter updates the system becomes:

\begin{equation}
    \begin{bmatrix}
        \mathbf{\dot{x}} \\
        \mathbf{\dot{\Hat{\theta}}}
    \end{bmatrix} = 
    \begin{bmatrix}
        \mathbf{f}(\mathbf{x}) + \mathbf{F}(\mathbf{x})\mathbf{\theta^{*}} + \mathbf{g}(\mathbf{x})\mathbf{u} \\
        \mathbf{\Gamma} \mathbf{\tau}(\mathbf{x}, \mathbf{\Hat{\theta}})
    \end{bmatrix}
\end{equation}

Solutions to this system evolve in $\mathcal{X} \times \Theta$. We can design an adaptively stabilizing controller by the notion of Lypunov stability, called \textit{adaptive Control Lyapunov Function (aCLF)}.

$V_{a}$ is a CLF for the system,

\begin{equation}
        \mathbf{\dot{x}} = \mathbf{f}(\mathbf{x}) + \mathbf{F}(\mathbf{x})\mathbf{\lambda}_{clf}(\mathbf{x}, \mathbf{\theta}) + \mathbf{g}(\mathbf{x})\mathbf{u},
\end{equation}

where,

\begin{equation}
    \mathbf{\lambda}_{clf}(\mathbf{x}, \mathbf{\theta}) \triangleq \mathbf{\theta} + \mathbf{\Gamma}(\frac{\partial V_{a}}{\partial \mathbf{\theta}}(\mathbf{x}, \mathbf{\theta}))^{T}
\end{equation}

aCLF constraint becomes,

\begin{align}
        inf_{u \in \mathcal{U}}[\frac{\partial V_{a}}{\partial \mathbf{\theta}}(\mathbf{f}(\mathbf{x}) + \mathbf{F}(\mathbf{x})\mathbf{\lambda}_{clf}(\mathbf{x}, \mathbf{\theta}) + \mathbf{g}(\mathbf{x})\mathbf{u})] \\ \leq -\alpha_{3}(\| \mathbf{x} \|, \mathbf{\theta})
\end{align}

We can extend this definition of CLF to the context of safety and define adaptive control barrier function. For the system (19) we first extend the definition of safe set (7)-(9) to be parameter dependent.

\begin{align}
    \begin{split}
        \mathcal{C}_{\mathbf{\theta}} \triangleq{}& \{\mathbf{x} \in \mathcal{X} : h_{a}(\mathbf{x}, \mathbf{\theta}) \geq 0 \}
    \end{split} \\
    \begin{split}
        \partial\mathcal{C}_{\mathbf{\theta}} \triangleq{}& \{\mathbf{x} \in \mathcal{X} : h_{a}(\mathbf{x}, \mathbf{\theta}) = 0 \}
    \end{split} \\
    \begin{split}
        Int(\mathcal{C}_{\mathbf{\theta}}) \triangleq{}& \{\mathbf{x} \in \mathcal{X} : h_{a}(\mathbf{x}, \mathbf{\theta}) > 0 \}
    \end{split}
\end{align}

{\bf Definition 3:} Let $\mathcal{C}_{\mathbf{\theta}} \subset \mathcal{X}$ be a family of 0-superlevel sets containing continuously differentiable function $h_{a} : \mathcal{X} \times \Theta \to \mathbb{R}$, with $\frac{\partial h_{a}}{\partial \mathbf{x}}$ Lipschitz continuous. Then $h_{a}$ is an \textit{adaptive Control Barrier Function (aCBF)} on the family of sets $\mathcal{C}_{\mathbf{\theta}}$ with adaptive gains $G$ for any $\theta \in \Theta$ and $\mathbf{\Gamma} \in G$ \cite{taylor2020adaptive}:

\begin{equation}
    sup_{u \in \mathcal{U}}[\frac{\partial h_{a}}{\partial \mathbf{\theta}}(\mathbf{f}(\mathbf{x}) + \mathbf{F}(\mathbf{x})\mathbf{\lambda}_{cbf}(\mathbf{x}, \mathbf{\theta}) + \mathbf{g}(\mathbf{x})\mathbf{u})] \geq 0)
\end{equation}

with,

\begin{equation}
        \mathbf{\lambda}_{cbf}(\mathbf{x}, \mathbf{\theta}) \triangleq \mathbf{\theta} + \mathbf{\Gamma}(\frac{\partial h_{a}}{\partial \mathbf{\theta}}(\mathbf{x}, \mathbf{\theta}))^{T}
\end{equation}

Similar to other CBF quadratic program-based controllers, we can formulate an aCBF-QP controller that filters a desired control input $\mathbf{k}_{d} : \mathcal{X} \times \Theta \to \mathcal{U}$ to drive the system to safety \cite{taylor2020adaptive}:

\begin{align}
        \mathbf{k}(\mathbf{x}, \mathbf{\Hat{\theta}}) = \argmin_{u \in \mathcal{U}} \| u - \mathbf{k}_{d}(\mathbf{x}, \mathbf{\Hat{\theta}})\|^{2} \\
    \textrm{s.t.} \ (30)
\end{align}

Similar to other CBF-QPs and aCLF-QP this also has a closed form solution.

The above shown aCBF is a basic framework to incorporate parameter uncertainties. Numerous other research studies have been conducted which showcases a more robust aCBF for varying parameter and model uncertainties. The state of the art in aCBFs include auxilary-variable aCBFs which improves feasibility of the CBF-QP while avoiding extensive parameter tuning with non-overshooting control \cite{liu2023auxiliary}, data-driven High Order Robust aCBF \cite{cohen2022high,lopez2020robust} which reduce parameter uncertainties in the estimates via online data recording, and learning-based aCBFs for unknown dynamics and model uncertainties \cite{xiao2023safe}. 

\section{Applications}

Barrier functions in general have had applications in various fields. Originally they have been used extensively in mathematics, like in optimization to keep the solution from going into unwanted regions to a much modern use case in reinforcement learning. In a similar fashion concept of barrier certificates has been previously utilized in verifying the safety of nonlinear and hybrid systems \cite{ames2019control,ames2016control,xiao2023safe}.

From our previous discussion of control barrier functions, we have laid down a framework for utilizing barrier function-based safety guarantees in modern control systems with autonomous characteristics. Here we are going to discuss some of these applications that have been implemented with varying degrees of success.

\subsection{Adaptive Cruise Control (ACC)}

CBFs have been extensively studied for use in the automotive domain. Modern automotive systems are characterized by their increasing complexity, integrating advanced technologies to enhance performance, efficiency, and safety. As vehicles become more sophisticated, ensuring compliance with safety-critical constraints becomes paramount. Advanced Driver Assistance Systems (ADAS) is an example of safety-critical constraints in modern vehicles. Control Barrier Functions (CBFs) have proven to be invaluable in addressing these challenges, offering a robust framework to enforce constraints and guarantee the safety and stability of automotive systems.

Systems like, adaptive cruise control, collision avoidance, lane keep assist, and tire slip control are some examples where numerous safety-critical constraints need to be satisfied for safe operation of a vehicle.

\textit{Adaptive Cruise Control (ACC)} has been the pinnacle of technology for semi-autonomous driving in modern vehicles. In ACC, a user is able to set a desired speed for the vehicle, when there are no vehicles immediately in front in the lane. When there is a vehicle in front the system tries to maintain a safe distance while following the vehicle in front.

We can setup the dynamics of the ACC problem as shown in \cite{ames2016control} by assuming a lead vehicle and a following vehicle as point masses. The following vehicle is equipped with ACC and has the objective of cruising at the desired speed provided by the user. The lead vehicle cruising with a constant velocity acts as a disturbance to this system which would reflect a real-world scenario.

The dynamics are:

\begin{equation}
    \dot{x} = \begin{bmatrix}
        v_{f} \\
        -F_{r}/M \\
        x_{2} - x_{1}
    \end{bmatrix} + \begin{bmatrix}
        1/M \\
        0 \\
        0
    \end{bmatrix} u
\end{equation}

Here, $v_{f}$ and $v_{l}$ are the velocities of the following vehicle and lead vehicle in $m/s$ respectively. $D$ is the distance between the two vehicles in $m$. $M$ is the mass of the following vehicle. $F_{r}(x) = f_{0} + f_{1}v_{f} + f_{2}v^{2}_{f}$ is the aerodynamic drag in N and $f_{0}, \ f_{1}$, and $f_{2}$ are the drag coefficients. $u \in U \subset \mathbb{R}$ is the control input for the following vehicle in N. For this particular problem, we consider the input to be constrained as $-mc_{d}g \leq u \leq mc_{a}g$. We express the safety constraint as

\begin{equation}
    D - \tau_{d}v_{f} \geq 0,
\end{equation}

for the dynamics (34), and we write the candidate CBF as follows,

\begin{equation}
    h(x) = D - \tau_{d}v_{f} - \frac{1}{2c_{d}g}(v_{f} - v_{l})^{2}
\end{equation}

Since, $h(x) > 0$ for any $x \in Int(\mathcal{C})$ also $L_{g}h(x > 0)$ which implies that $h$ has a relative degree of 1. The CBF-CLF constraint becomes,

\begin{equation}
    sup_{u \in U}[L_{f}h(x) + L_{g}h(x)u] \geq - \alpha (h(x)),
\end{equation}

\begin{equation}
        sup_{u \in U}[L_{f}V(x) + L_{g}V(x)u] \leq - \gamma (V(x)),
\end{equation}

\begin{figure}[!t]
\centerline{\includegraphics[width=1\columnwidth]{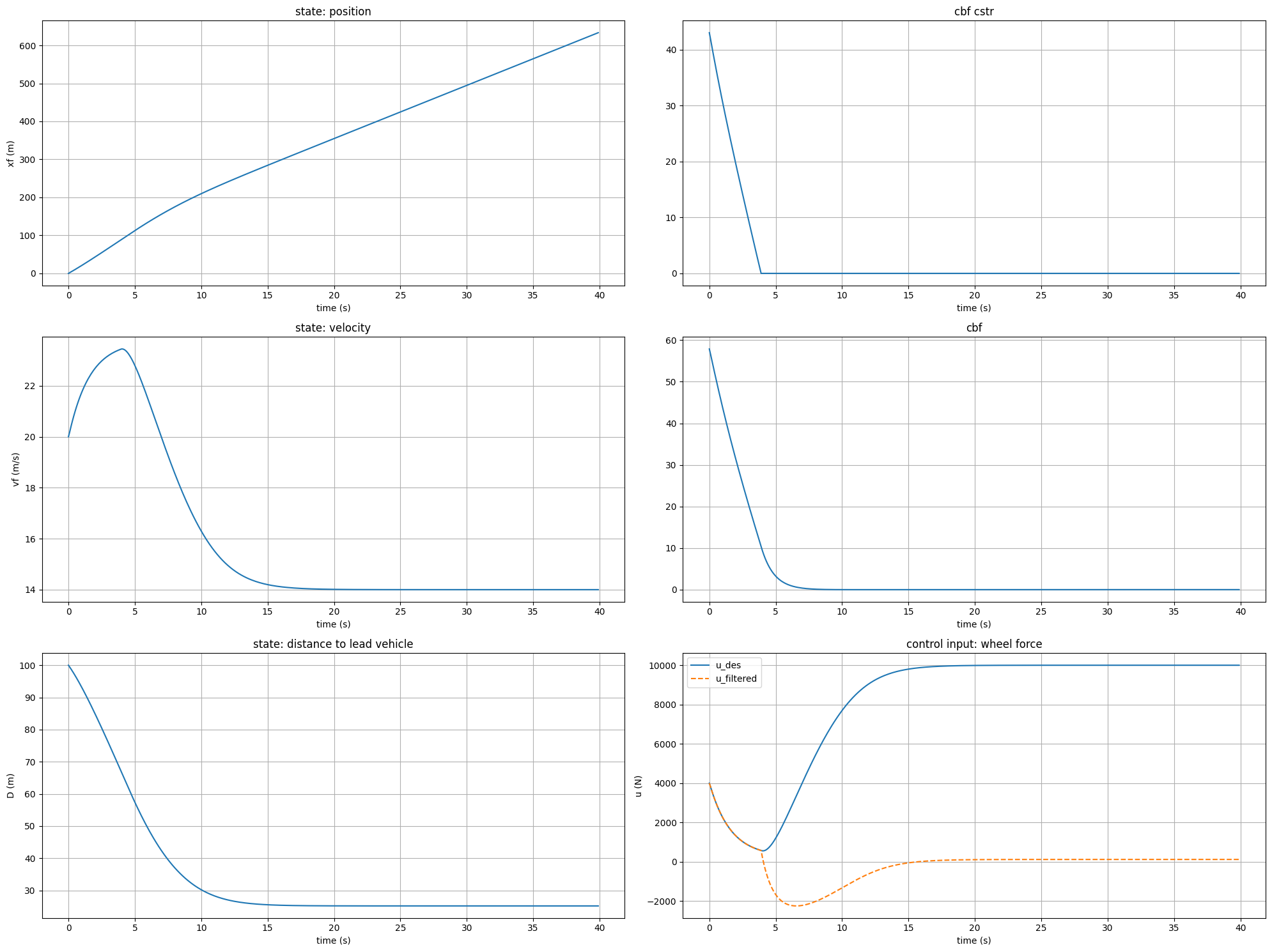}}
\vspace{-2mm}
\caption{ACC Simulation Results}
\label{fig:cbf-acc}
\vspace{-6mm}
\end{figure}

Now we can use the above CBF-CLF constraint to formulate the CBF-CLF-QP. The CBF-CLF-QP would filter a desired control input to make sure the system attains stability and remains inside the safe set. The results shown in Fig. ~\ref{fig:cbf-acc} were obtained after simulating the above system from time $t:$ 0 to 40 seconds, with initial conditions as $x_{f} = 0$, $v_{f} = 20 m/s$, $v_{l} = 14 m/s$, $v_{d} = 24 m/s$ and $D = 100 m$.

From the plots in Fig. ~\ref{fig:cbf-acc} we can see that in the first 5 seconds, the system tries to attain the desired velocity, but as the following vehicle gets closer to the lead vehicle, because of the CBF constraint the vehicle starts decelerating and after 15 seconds stabilizes to maintain a constant velocity matching the velocity of the lead vehicle.

The system shown above is a simplified version of ACC which has relative degree 1. But modeling the system with more realistic constraints would cause the system to become high order. In \cite{xiao2023safe,xiao2019control,cohen2022high} have shown the high order ACC problem can be solved using HOCBFs. Moreover, an ACC problem with parameter uncertainties has been successfully solved using aCBFs as shown in \cite{cohen2022high,xiao2023safe,xiao2019control,taylor2020adaptive}.

\cite{xiao2023safe} also shows several scenarios where HOCBFs and aCBFs are utilized for traffic management of autonomous vehicles. From highway lane merging to round-about scenarios, CBFs have been shown to work successfully ensuring safety. 

\subsection{Robotic Systems}

CBFs coupled with CLFs have been successfully utilized in legged robotic systems. \cite{ames2019control} showcases how CBF-CLF-QP with actuator constraints can be utilized for implementing dynamic walking in bipedal robot over a terrain of stepping stones with variable step lengths. The results showcase that due to CBF being enforced the robot maintains safety.

\cite{xiao2023safe} shows multiple scenarios where CBF is implemented in ground robots for collision avoidance. Multiple ground robots are given the objective to reach a target point while avoiding an obstacle in the way. CBF-CLF-QP achieves stability by reaching the target point and also maintaining safety by avoiding collision with the obstacle.

Learning-based aCBFs are shown in \cite{xiao2023safe}, which are utilized in ground robots for safe exploration in unknown environments. The controller avoids collision with obstacles without prior knowledge of their position and maintains safety.

Another system where CBFs can be used to maintain safety is the unicycle system. \cite{ames2019control,xiao2023safe} show a similar system of a single-wheeled robot that is prevented from falling by the use of CBFs.

\subsection{Quadrotor UAVs}

Quadrotor UAVs have risen in popularity because of their dynamic nature that allows them to be deployed in varied scenarios, from delivering goods to surveillance in indoor spaces. Due to this dynamic nature, they can be difficult to control in safety-critical situations. Collision avoidance is a paramount concern, as it directly influences the safety and effectiveness of the UAVs deployed in such an environment. In cooperative and coordinated autonomous operations with multiple UAVs, collisions can be controlled and avoided with more certainty, as we have full access to the control parameters of all the UAVs and can perform deterministic operations, e.g., coordinated UAV swarm, platooning etc.

CBFs have been deployed with varying success for safe operations of quadrotors. \cite{tayal2023control,xu2018safe,xiao2023safe} showcase successfully deployed CBF-based controllers for collision avoidance in quadrotors. In these experiments, the CBF-QP is generally paired as a filter for a trajectory-tracking controller.

\cite{lopez2020robust} shows a simulation study of an aircraft's attitude control with model uncertainties. A data-driven approach is showcased for designing robust aCBF, that successfully stabilizes and guarantees safety for the system.

\section{Conclusion}

This survey paper has showcased the intricate landscape of Control Barrier Functions (CBFs) and their extensions, providing a comprehensive understanding of their theoretical foundations, practical applications, and the evolving field of research. From the fundamental principles of CBFs to the sophistication of High Order Control Barrier Functions (HOCBFs), and the adaptive nature of Control Barrier Functions (aCBFs), our exploration has showcased the versatility of these frameworks in addressing safety and stability challenges across diverse domains.

The applications of CBFs in automotive systems, as outlined in Section III, exemplify their pivotal role in ensuring the safety of modern vehicles. From collision avoidance to adaptive cruise control and compliance with road regulations, CBFs offer a flexible and reliable approach to navigating the intricate dynamics of the automotive environment.  The adaptability of Control Barrier Functions becomes particularly crucial in the face of uncertainties and evolving operating conditions, this is further exemplified by their use in highly dynamic quadrotor UAVs, highlighting their relevance in real-world, dynamic scenarios. 

\bibliographystyle{ieeetr}
\bibliography{reference}

\end{document}